\documentclass[%
superscriptaddress,
preprint,
nofootinbib,
amsmath,amssymb,
aip,
]{revtex4-2}
\usepackage{graphicx}
\usepackage{xcolor}
\usepackage{media9}
\usepackage{float}
\usepackage{subcaption}
\usepackage{caption}
\usepackage{lipsum} % <- For dummy text

\def\vec{\mathbf}
\def\t#1{\text{#1}}

\begin{document}
\title{Inverse-designed Metastructures together with Reconfigurable Couplers to Compute Forward Scattering}
\author{Vahid Nikkhah}
\thanks{These authors contributed equally}
\affiliation{University of Pennsylvania, Department of Electrical and Systems Engineering, Philadelphia, PA 19104 U.S.A.}

\author{Dimitrios C. Tzarouchis} 
\thanks{These authors contributed equally}
\affiliation{University of Pennsylvania, Department of Electrical and Systems Engineering, Philadelphia, PA 19104 U.S.A.}

\author{Ahmad Hoorfar} 
\affiliation{Villanova University, Department of Electrical and Computer Engineering, Villanova, PA 19085 U.S.A.}

\author{Nader Engheta}
\thanks{Corresponding author: engheta@seas.upenn.edu}
\affiliation{University of Pennsylvania, Department of Electrical and Systems Engineering, Philadelphia, PA 19104 U.S.A.}

% \date{\today}

%--------------------------------------------------
\begin{abstract}
    Wave-based analog computing in the forms of inverse-designed metastructures and the meshes of Mach--Zehnder interferometers (MZI) have recently received considerable attention due to their capability in emulating linear operators, performing vector-matrix multiplication, inverting matrices, and solving integral and differential equations, via electromagnetic wave interaction and manipulation in such structures. Here, we combine these two platforms to propose a wave-based metadevice that can compute scattered fields in electromagnetic forward scattering problems. The proposed device consists of two sub-systems: a set of reconfigurable couplers with a proper feedback system and an inverse-designed inhomogeneous material block. The first sub-system computes the magnitude and phase of the dipole polarization induced in the scatterers when illuminated with a given incident wave (matrix inversion). The second sub-system computes the magnitude and phase of the scattered fields at given detection points (vector-matrix multiplication). We discuss the functionality of this metadevice, and through several examples, we theoretically evaluate its performance by comparing the simulation results of this device with full-wave numerical simulations and numerically evaluated matrix inversion. We also highlight that since the first section is reconfigurable, the proposed device can be used for different permittivity distributions of the scatterer and different incident excitations without changing the inverse-designed section. Our proposed device may provide a versatile platform for rapid computation in various scattering scenarios. 
\end{abstract}

\maketitle

%--------------------------------------------------
\section{Introduction}

Optical analog computing shows great promise for next-generation computing platforms. Owing to some of the features of light-matter interaction, optical metastructures gives way for low-power, and high-speed analog computations~\cite{miller_2010,miller_2017,caulfield_dolev_2010,bogaerts_perez_capmany_miller_poon_englund_morichetti_melloni_2020,Zangeneh-Nejad2021}. The artificially-engineered electromagnetic structures known as metamaterials~\cite{engheta_ziolkowski_2006} combined with inverse-design techniques enable engineering of wave propagation through material structures down to subwavelength scales with unprecedented efficiency~\cite{Piggott2015,hughes_minkov_williamson_fan_2018}. These structures are capable of doing computation with near the speed of light for complex computational problems such as solving integral equations~\cite{Estakhri2019}. Moreover, the wave propagation through material structures exhibits parallelism in terms of encoding and processing information~\cite{camacho_edwards_engheta_2021}. Considering the capabilities of optical analog computing, here we propose the idea of wave-based metadevices for performing specialized computational tasks, particularly for certain physical problems such as the electromagnetic forward-scattering problems, which can be computationally very intensive.

In this work, we encode the information about the quantities of interest, i.e., the complex-valued electric field and induced polarization in our scattering problem, into the complex amplitudes of the waveguide modes. Then we process them by a metadevice, schematically shown in Fig.~(\ref{Idea_figure}), that consists of two sub-systems: (a) a set of reconfigurable couplers, e.g., Mach--Zehnder interferometers (MZIs), and (b) an inverse-designed metastructure. In the first part, a network of reconfigurable couplers is designed, which when endowed with a proper feedback loop, can compute the inverse of a given matrix. This part is designed to compute the magnitude and phase of the dipole polarization induced inside the target scatterers by the local electric field, for a given incident field. As discussed below, since these couplers are tunable the permittivity of implemented scatterers and the incident fields can be modified at will. The second part is an inverse-designed metastructure that can perform a matrix-vector product with waves for computing the scattered fields, generated by the polarization induced in the target, at the designated detection points
using the Green's function as the propagator. This part of our metadevice is formed using an inhomogeneous distribution of dielectric materials, whose optimized distribution is obtained using the method of inverse-design. Finally, we validate the performance of the proposed design through simulations of a series of forward scattering examples. 

The article is organized as follows: In section II, a brief overview of the mathematical formalism for solving forward scattering problems using the method of discrete dipole approximation (DDA) is given. In section III, the details of the metadevice design are presented, i.e., the tunable couplers and the inverse-designed stages. We then test, via simulation, the proposed device against two particular examples of a forward scattering problem in section IV, i.e., (a) an example of a scattering problem with various incident fields and various permittivity profiles and (b) an example of a through-the-wall radar imaging problem. The summary and the concluding remarks are given in section V.  

%--------------------------------------------------
\section{Formulation Overview}
\begin{figure}[h]
    \centering
   \includegraphics[width = 1\textwidth]{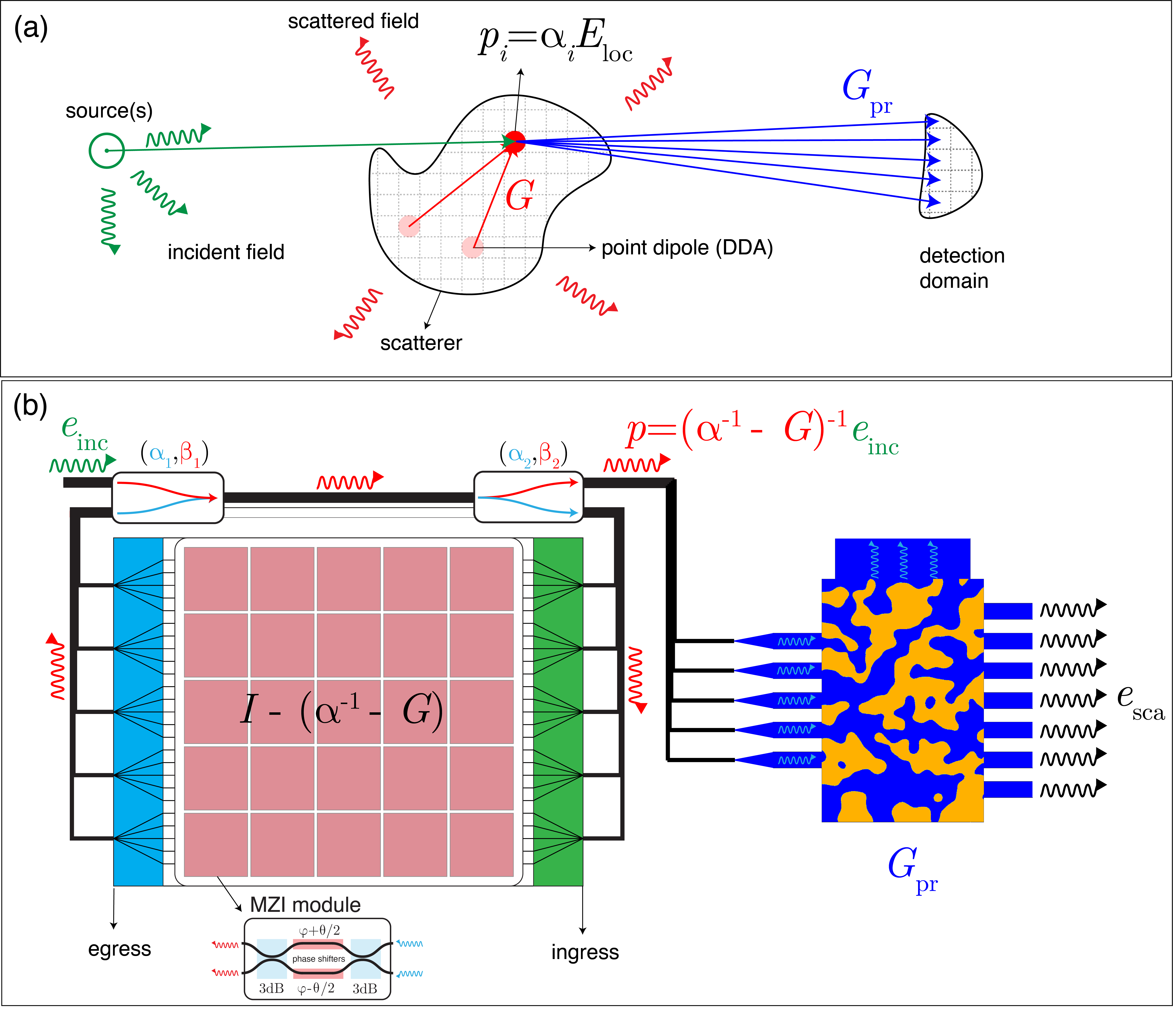}
   \caption{{\bf Proposed metadevice that emulates the forward scattering problem}: (a) the physical problem that consists of the incident excitation (green arrows), the scattered field (red) and the scattered field obtained at a given set of detection points. Computing the scattered fields requires the matrix inversion for the determination of the induced dipole polarization vectors at the scatters due to the incident excitation, and a matrix-vector operation for the evaluation of the scattered field at the desired detection domain. (b) Our proposed device consists of the reconfigurable direct-complex-matric (DCM) architecture involving reconfigurable multipliers (for matrix inversion) and the inverse-designed metastructure (for vector-matrix multiplication). The first emulates the interaction between the induced dipoles of the scatterers (following DDA), whereas the second evaluates the propagator.}
   \label{Idea_figure}
\end{figure}

We start with the standard integral representation of the electromagnetic scattering for a two-dimensional (2D) problem: \begin{equation}\label{Integral_Eq}
    \vec{E}(\vec{r})=\vec{E}_\t{inc}(\vec{r})+\int_S\vec{G}(\vec{r},\vec{r}^\prime)\chi(\vec{r}^\prime)\vec{E}(\vec{r}^\prime)d^2\vec{r}^\prime
\end{equation}
where $\vec{E}(\vec{r})$ and  $\vec{E}_\t{inc}(\vec{r})$ are, respectively, the total  and the incident electric fields at the observation point $\vec{r}$, $\vec{G}(\vec{r},\vec{r}^\prime)$ is the 2D Green's function of the Helmholtz’s equation and $\chi(\vec{r}^\prime)=\varepsilon_0(\varepsilon_{r}(\vec{r}^\prime)-1)$ is the susceptibility (sometimes referred to as contrast or scattering potential) of the object with relative permittivity $\varepsilon_{r}(\vec{r}^\prime)$ and enclosed by the cross-sectional surface $S$ as shown in Fig.~(\ref{Idea_figure}) (a). Here we have assumed that the surrounding medium is air with permittivity $\varepsilon_0$. Indeed, Eq.~(\ref{Integral_Eq}) is a Fredholm integral equation of the 2nd kind that is often referred to as the Lippmann--Schwinger equation~\cite{colton_kress_2019} and lies within the heart of many numerical electromagnetic methods, such as the discrete dipole approximation (DDA)~\cite{Purcell1973,Draine1994}, the contrast source inversion method (CSI)~\cite{Wang1989,Berg1997}, the coupled dipole method~\cite{Lakhtakia1990}, and the Method of Moments (MoM)~\cite{Kahnert2003}. 

We need to express Eq.~(\ref{Integral_Eq}) in terms of its equivalent discrete version using matrix/vector quantities. For this purpose we closely follow the DDA approach, a general method for solving electromagnetic scattering problems of dielectric targets of arbitrary shapes and compositions~\cite{Yurkin2007}. The domain $S$ that contains the 2D dielectric object is discretized in small scattering cells $i=1,2,3,\dots N$, where each 2D cell exhibits an induced dipole moment $\vec{p}_i$ that is generally proportional to the geometrical shape (here surface) and the associated permittivity at the $i$-th cell (assuming piecewise homogeneous scatterers) and the local electric field (Fig.~(\ref{Idea_figure}) (a)). The evaluation of the unknown induced dipole moments is the first goal of our approach. 
Each cell acquires its dipole moment due to the local electric field, i.e.,
\begin{equation}\label{Pol_Efield}
    \vec{p}_i=\alpha_i\vec{E}^\t{loc}_i
\end{equation}
where the $\vec{E}_i^\t{loc}$ is the local electric field at the center of the cell $i$ and $\alpha_i$ is the polarizability that depends on the shape and the material composition of each 2D cell. In turn, the local field is generated by the incident field and the secondary fields generated from all the rest of the dipoles (other discrete cells) such that:  
\begin{equation}\label{localEfield_Pol}
    \vec{E}_i^\t{loc}=\vec{E}_i^\t{inc}+\sum_{k\neq i}^N\vec{G}_{ik}\vec{p}_{k}
\end{equation}
where $\vec{E}^\text{inc}_i$ is the incident field at the target point $i$, $\vec{p}_k$
is the polarization induced inside the target cell $k$
and $\vec{G}_{i k}$ is the 2D Green's function for calculating the fields, generated from $\vec{p}_k$,
at the location of the target point $i$. In our case we consider a two-dimensional (2D) problem with a transverse electric (TE) excitation. Therefore the corresponding Green's function reads \begin{equation}\label{hankelGreen}
    \vec{G}_{ik} = \vec{G}(\vec{r}_i-\vec{r}_k)=-j\frac{k_0^2}{4\pi \varepsilon_0}H^{(2)}_0(k_0|\vec{r}_i-\vec{r}_k|)
\end{equation}
where $H^{(2)}_0(k_0|\vec{r}_i-\vec{r}_k|)$ is Hankel function of the second type and 0-th order and $k_0 = \omega_0 \sqrt{\mu_0 \varepsilon_0}$ is the free-space wavenumber ~\cite{balanis} (here we use the $e^{j\omega t}$ convention). By combining Eqs.~(\ref{Pol_Efield}) and~(\ref{localEfield_Pol}) we get the following expression:
\begin{equation}
    \alpha^{-1}_i\vec{p}_i=\vec{E}^\t{inc}_i+\sum_{i\neq k}^N\vec{G}_{ik}\vec{p}_{k}
\end{equation}
for which the unknown quantity is the dipole polarization induced inside the cells, i.e. $\vec{p}_i,\quad i=1,..,N$ . Using simple algebra, the above system can be arranged using the matrix formulation as follows
\begin{equation}\label{pol_Einc}
    p=\left(\t{diag}(\alpha^{-1})-G\right)^{-1}e_\t{inc}
\end{equation}
where the lowercase quantities $p=[\vec{p}_1,\vec{p}_2,...,\vec{p}_N]^T, \alpha=A_{\text{cell}}\varepsilon_0[\varepsilon_1-1,\varepsilon_2-1,...,\varepsilon_N-1]$ ($A_{\text{cell}}$ is the cross-sectional area of a cell) and $e_\t{inc}=[\vec{E}^\t{inc}_1,\vec{E}^\t{inc}_2,...,\vec{E}^\t{inc}_N]^T$ are $\mathbb{C}^{N\times1}$vectors, $\t{diag}(\cdot)$ is the diagonal matrix operator, and $G$ is a $\mathbb{C}^{N\times N}$ Toeplitz matrix with zero diagonal entries (assuming a uniformly spaced discrete grid)~\cite{Groth2020}. 

Finally, the scattered field observed at $M$ specified discrete detection points (in general $M\neq N$) is given by: 
\begin{equation}\label{Escat_Einc}
    e_\t{sca}=G^\t{pr}~p=G^\t{pr}\left(\t{diag}(\alpha^{-1})-G\right)^{-1}e_\t{inc}
\end{equation}
where $G^{pr}\in\mathbb{C}^{M\times N}$ is the ``propagator" Green's function matrix connecting the induced dipole polarization vectors of subscatterers (cells) and the detection points. Based on the above matrix representations of the scattering problem we propose a specially designed wave-based reconfigurable metadevice that is able to simultaneously solve Eq.~(\ref{pol_Einc}) and implement the matrix-vector operation of Eq.~(\ref{Escat_Einc}) for different excitations and for different scattering scenarios.

\section{Metadevice design}
In this section we present the main results of our study. As sketched in Fig.~(\ref{Idea_figure}) (b), our proposed metadevice consists of two parts: (A) the matrix inversion part and (B) the vector-matrix multiplication part. For the first part we utilize a recently introduced Direct Complex Matrix (DCM) reconfigurable architecture~\cite{Tzarouchis2021} properly connected to a feedback loop, capable of performing the desired matrix inversion in Eq.~(\ref{pol_Einc}). For the second part we propose an inverse-designed metastructure that is able to conduct the matrix-vector product for emulating the propagator matrix, $G^\t{pr}$ in Eq.~(\ref{Escat_Einc}). 

\subsection{Reconfigurable DCM network stage for inverting matrices}
% Here we descibe briefly the DCM architecture. 
This stage, shown as the left panel of Fig.~(\ref{Idea_figure}) (b), follows the architecture introduced in~\cite{Tzarouchis2021}. In particular, a given $N\times N$ complex-valued matrix can be implemented in this device for inversion that has three sections: (a) the ingress, (b) the middle, and (c) the egress sections. The ingress section consists of $N$ number of $1\to N$ signal splitters, each of which splits each of the $N$ inputs into $N$ outputs. All such $N^2$ outputs of the ingress section are then properly routed to the middle section that consists of $N^2$ dedicated tunable components, each of which consists of a combination of a phase shifter and an amplifier/attenuator, which we call a ``multiplier".  These $N^2$ multipliers directly represent $N^2$ elements of the matrix in Eq.~(\ref{pol_Einc}) to be inverted.  Finally, the $N^2$ outputs of the middle sections are then routed to the egress section, i.e., $N$ number of $N\to1$ combiners\footnote{As an aside, it is worth pointing out that such combiners are not power combiners.  Instead, the output of each $N\to1$ combiner is proportional to the fraction of the algebraic sum of complex-valued amplitudes of $N$ input signals, and consequently the output power is usually less than the sum of the input power; As a result, the remaining power goes to some ``loss channel" e.g., either dissipates in the element, scattered away from the system, reflected, or goes out of the system in unused channels.  Consequently, we need to have amplifiers to bring the output signal level up to the desired value for the feedback system}.
As such the egress section is basically inverse of the egress section. Moreover, there is a feedback loop connecting the outputs to the inputs.  

For the feedback loop section we used two identical couplers that split the input and output signal with a coupling ratio $\beta_{1,2}$ (and the through is $|\alpha_{1,2}|^2=1-|\beta_{1,2}|^2$, see Fig.~(\ref{Idea_figure}) (b)) and the supporting information. The output of the DCM stage estimates a value that is proportional to the induced dipole polarization vector $p$. In order to retrieve the actuall value of the polarization vector we need (a) to scale the input vector $e_{inc}$, i.e., $e_\t{inc}/\beta_1$, and (b) scale the output, i.e., $p_\t{DCM}/\beta_2$. Therefore, from Eq.~(\ref{pol_Einc}) the total scaling factor $\frac{1}{\beta_1\beta_2}$ should be applied to the DCM output. Notice that in all examples presented below we assume couplers with $-30$dB coupling ($|\alpha_1|^2=|\alpha_2|^2=0.999$ and $|\beta_1|^2=|\beta_2|^2=0.001$) requiring a scaling factor of $\approx10^3$ to be applied at the output of this stage for the retrieval of the induced dipole polarization vector. (See the supplementary information for evaluation of dimensionless quantities in our analysis.)  

Overall, such device can implement the desired matrix inversion, required in Eq.~(\ref{pol_Einc})~\cite{Tzarouchis2021}.  Since the $N$ diagonal terms of this matrix depend on the relative permittivities of $N$ scattering cells, these diagonal modules of the middle section can be reconfigured by changing the phase and magnitude of their transfer functions (using the phase shifters and the amplifiers/attenuators in them) in order to accommodate different permittivities for these scatterers.  In short, when the complex values of the incident fields at the location of $N$ scatterers are inserted into the DCM stage of our metadevice, we evaluate the induced polarization vectors for different scattering scenarios represented by different material properties, e.g., permittivities, of the individual scatterers and different incident excitations.
\clearpage
\subsection{Inverse-designed metastructures for emulating propagator matrix}
\begin{figure}[h]
    \centering
    \includegraphics[width = \textwidth]{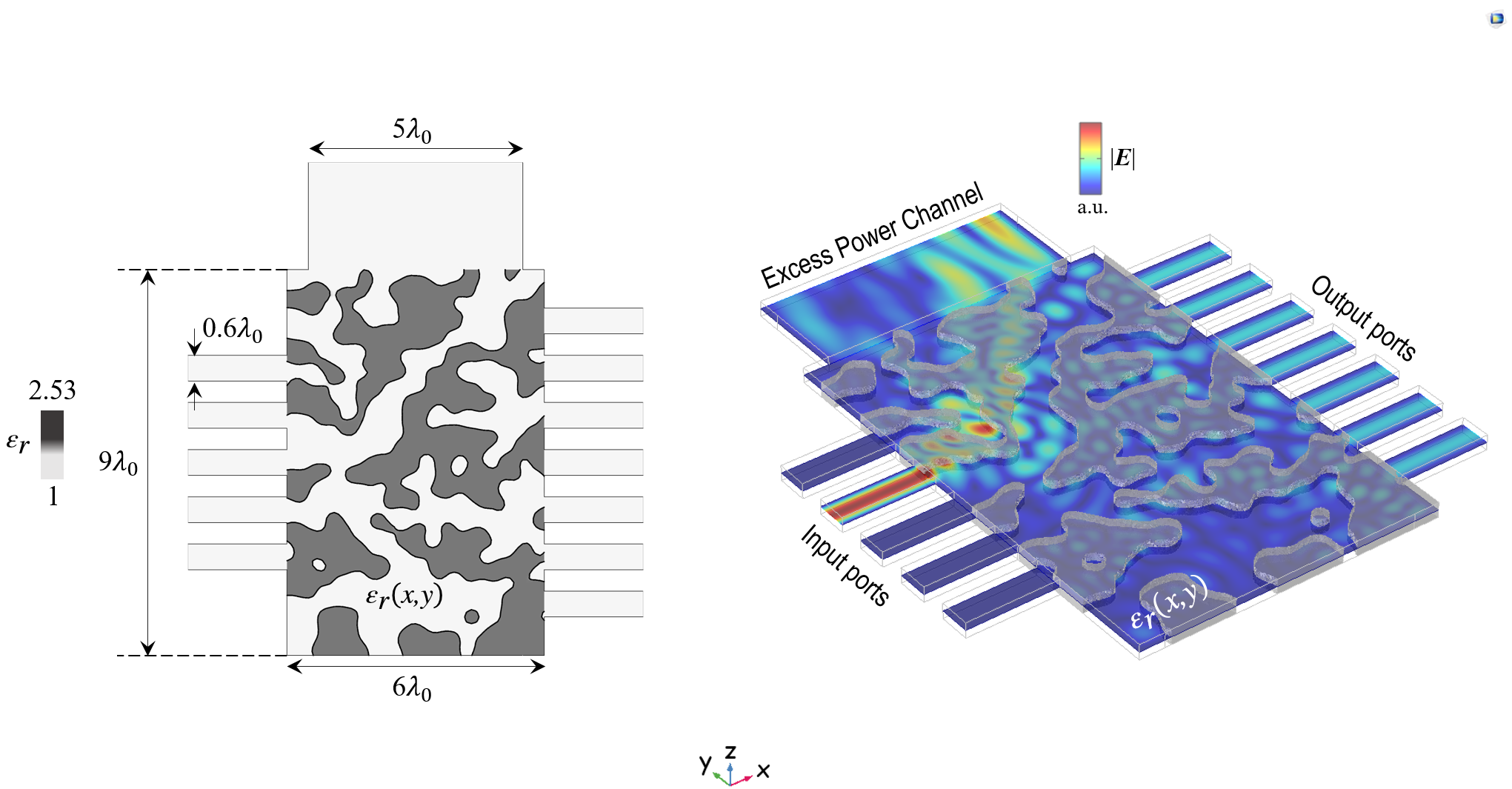}
    \caption{{\bf Inverse-designed metastructure}:  Inverse-designed structure designed for $f_0 = 3^\t{GHz}$  comprising 5 input and 7 output ports for emulating the propagator matrix, $G^\t{pr}$, of the configuration shown in Fig.~(\ref{mainResult}) (a). The size of the design region is $(L_x,L_y) = (6\lambda_0,9\lambda_0)$ and $0.3\lambda_0$ along $z$ direction. The material distribution is a function of $(x,y)$ only. The input and output ports are single mode $\t{TE}^{10}$ waveguides.  A multimode waveguide on top of the structure lets the excess power out (a) The optimized distribution of polystyrene with $\varepsilon_r = 2.53$ (dark grey regions) and air with $\varepsilon_r  = 1$ (light grey regions) (b) The magnitude of simulated electric field distribution inside the inverse-designed metastructure when the input port 2 is excited.}
    \label{Results_InverseDesign}
\end{figure}

In this section, we utilize the method of inverse design ~\cite{molesky_2018,bendsoe_sigmund_2019,piggott_lu_lagoudakis_petykiewicz_babinec_vuckovic_2015,piggott_petykiewicz_su_vuckovic_2017,sigmund_chigrin_2009,lu_boyd_vuckovic_2011,lalau-keraly_bhargava_miller_yablonovitch_2013}, which is a density-based topology optimization technique using a gradient-descent approach implemented by adjoint sensitivity analysis in COMSOL, to propose the wave-based inhomogeneous metastructures that emulate the propagator matrix $G^\t{pr}$ (see also supporting information). Figure \ref{Results_InverseDesign}(a) presents an example of such a structure that represents the propagator matrix for a 2D scattering example, shown in Fig.~(\ref{mainResult}) (a), with 5 scatterers and 7 detection points. This structure may consist of a metallic box with perfect electric conducting (PEC) walls, with 5 single-mode waveguides as the input ports and 7 single-mode waveguides as its output ports. The inputs are excited with the complex amplitudes of the induced dipole polarization vectors (which are the outputs of the reconfigurable DCM network), and the complex-valued scattered fields at the detection points can be obtained (after applying the scale factor) at the outputs of this metastructure,  according to $e_\t{sca}=G^\t{pr}p$ (see Fig.~(\ref{Results_InverseDesign}) (b)).

The size of the design domain is $6\lambda_0 \times 9\lambda_0 \times 0.3\lambda_0$ ($\lambda_0$ is the free space wavelength corresponding to the design frequency $f_0=3^{\t{GHz}}$). The size of the waveguides cross section is $0.6\lambda_0 \times 0.3\lambda_0$ supporting $\t{TE}^{10}$ propagating mode only. Since the $G^\t{pr}$ is not a unitary matrix, a waveguide on top of the metastructure takes the excess energy to outside. This waveguide has a wider width of $5\lambda_0$ to support multiple propagating modes for exiting the excess energy. The optimized material distribution inside the design region, for the example shown in Fig.~(\ref{mainResult}) (a), is shown in Fig.~(\ref{Results_InverseDesign}) (a). The dark grey regions are assumed to be polystyrene with $\varepsilon_{r,\t{poly}} = 2.53$ and the light grey regions are air. The material distribution is only a function of $(x,y)$ and it is invariant along the $z$ direction. Fig.~(\ref{Results_InverseDesign}) (b) shows the norm of the simulated electric field distribution inside the optimized structure when the input port 2 is excited, for the example shown in Fig.~(\ref{mainResult}) (a).

Since the elements of $G^\t{pr}$ only depend on the relative distances between locations of scattering cells and detection points, as long as the locations of the cells and the detection points are unchanged the designed metastructure can be used to compute the scattered fields for various forward scattering problems with different relative permittivity of the scatterers and/or various types of incident excitations.

In the next section, the simulation results for our proposed metadevice, which is the combination of the DCM stage and inverse-designed metastructure, are presented and evaluated for computing forward scattering in several examples. We show that the metadevice is capable of computing the complex amplitudes of the scattered fields at the detection points, for different incident radiations and various relative permittivities of the target scatterers.
\clearpage
\section{Examples and Discussion}
\begin{figure}[h]
    \centering
    \includegraphics[width = 1\textwidth]{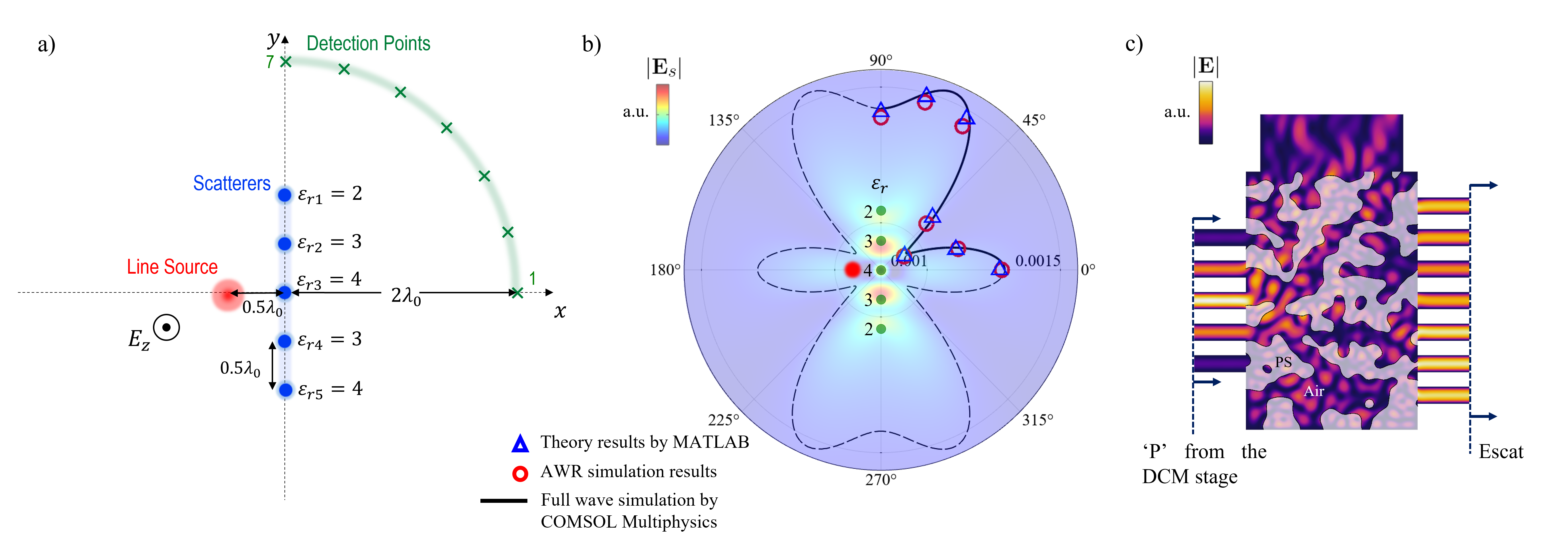}
    \caption{\textbf{Example 1 and its simulation results} (a) Geometry of the forward scattering problem under investigation as the first example. 5 two-dimensional (2) D dielectric rods, with subwavelength cross sections, are illuminated by an infinitely long line electric current source with the out-of-plane electric field. The scattered fields are obtained at 7 designated detection points on the circular path highlighted by green (b) The scattered field distribution computed by full-wave simulation using the COMSOL Multiphysics\textsuperscript{\textregistered}. The scattered field data on the detection path obtained by the full-wave simulation (black solid curve), from the DDA calculation in Matlab\textsuperscript{\textregistered} (blue triangles), and from the simulated outputs of our proposed metadevice (with AWR Microwave Office, red circles) are compared on a polar plot (c) Inverse-designed metastructure for emulating the propagator Green function. The inputs to this structure are excited with the complex amplitudes proportional to the p vector obtained by the DCM architecture. The mode amplitudes at the outputs are proportional to the scattered field complex amplitudes at the detection points}
    \label{mainResult}
\end{figure}

\begin{figure}[h]
    \centering
    \includegraphics[width = 1\textwidth]{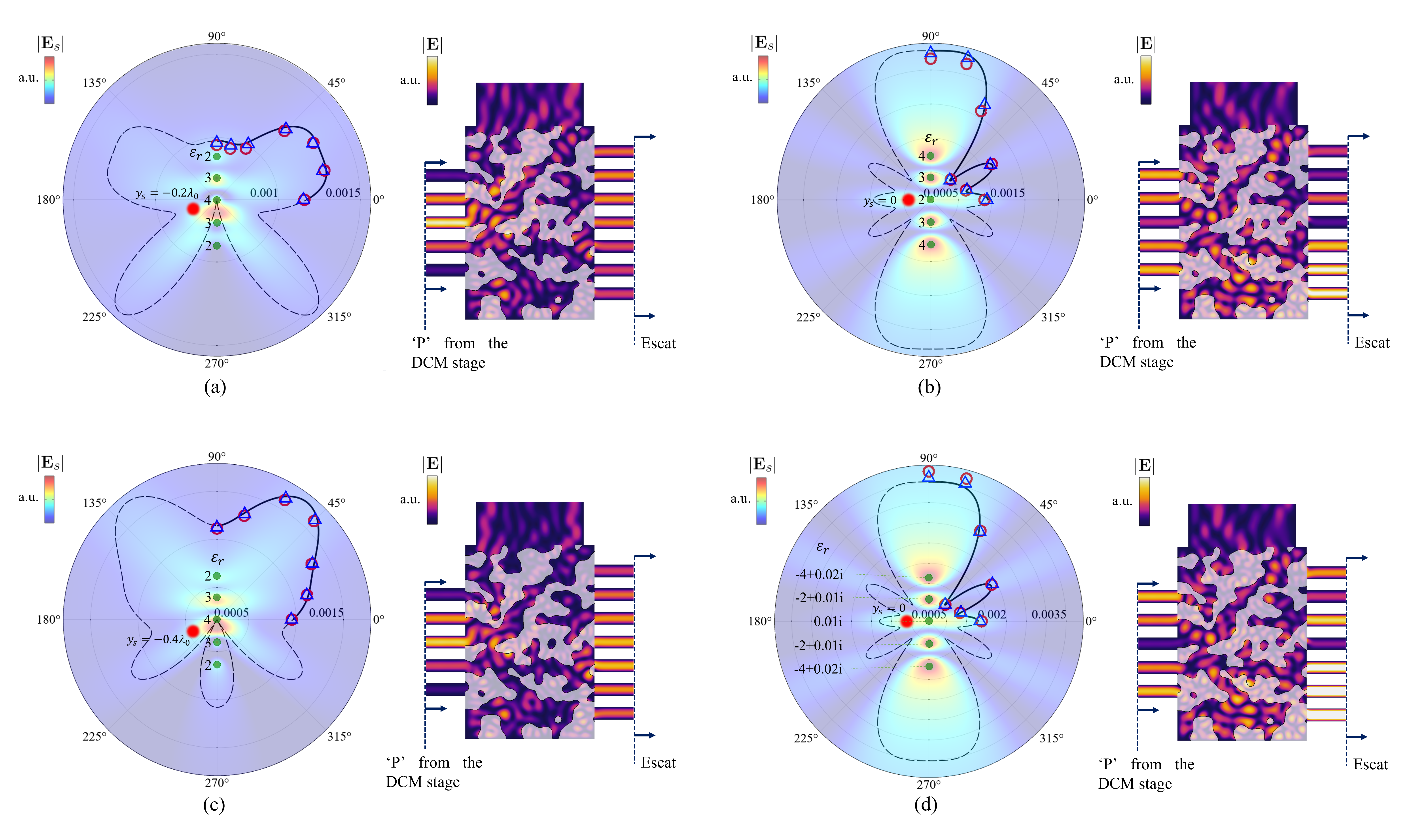}
    \caption{\textbf{Examples 2 (first row) and 3 (second row): Different incident excitation and different sets of relative permittivities of the rods.} Similar to the example of Fig.~(\ref{mainResult}), here we have the simulations results for the scattered fields (blue triangle Matlab\textsuperscript{\textregistered} results, red circle AWR Microwave Office\textsuperscript{\textregistered} simulation results, black line full wave simulations with COMSOL Multiphysics\textsuperscript{\textregistered}) for different incident radiations produced by a line source that is displaced by (a) $0.2\lambda_0$ and (b) $0.4\lambda_0$ downward relative to the center of the array. The second row shows the corresponding results for the modified relative permittivities for (c)  $\varepsilon_{r} = \{4,3,2,3,4\}$ and for (d) $\varepsilon_{r} = \{-4+j0.02,\ -2+j0.01,\ j0.01,\ -2+j0.01,\ -4+j0.02\}$.}
    \label{variousResults_epsilon_Einc}
\end{figure}

As the first example, depicted in Fig.~(\ref{mainResult}) (a), we consider the 2D scattering from 5 infinitely long cylindrical dielectric rods (parallel with the $z$ axis) that are separated by $\lambda_0 / 2$, placed along a linear grid on the $y$ axis. The diameter of each rod is chosen to be deeply subwavelength, $d = \lambda_0 /50$, to model each rod with a polarizability of $\alpha_i=A_{\text{rod}}\varepsilon_0(\varepsilon_i-1)$ where $A_{\text{rod}} = \pi d^2/4$ is the cross-sectional area of the rod and $\varepsilon_i$ is the relative permittvity of the rod $i,\quad i =1,2,...,5$. In this example, the relative permittivities of the rods are chosen to be $\varepsilon_{r} = \{2,3,4,3,2\}$ as shown in the figure. An infinitely long z-oriented electric line source, located $\lambda_0 /2$ away from the center of the array, is generating the incident electromagnetic field with an out-of-plane E-field polarization. The scattered fields generated from these 5 rods are to be evaluated at 7 designated detection points positioned on a quarter section of a circle with radius of $2\lambda_0$ as shown in Fig.~(\ref{mainResult}) (a). As we discuss below our proposed metadevice can evaluate the scattered fields from this structure, evaluated at the detection points. The DCM stage of the metadevice is designed to allow the inversion of the matrix containing the information about the relative permittivities and locations of these 5 scatterers (see Eq.~(\ref{pol_Einc})), and the inverse-designed metastructure section is designed to emulate the propagator part of Eq.~(\ref{Escat_Einc}) for this scattering problem.  Fig.~(\ref{mainResult}) (c) shows this inverse-designed metastructure.  The DCM stage is simulated using the AWR Microwave Office\textsuperscript{\textregistered} and the wave interaction in the inverse-designed metastructure is simulated using the COMSOL Multiphysics\textsuperscript{\textregistered} ~\cite{COMSOL} (see the supporting information). As depicted in Fig.~\ref{Idea_figure} (b), the complex-valued incident electric fields at the locations of 5 rods are considered as the inputs to the DCM stage, and the complex-valued induced dipole polarization amplitudes excited from the DCM stage are the inputs to the inverse-designed metastructures.  Fig.~(\ref{mainResult}) (b) depicts the magnitude of the scattered electric field, obtained from full-wave simulations using the COMSOL Multiphysics\textsuperscript{\textregistered}~\cite{COMSOL}.  The solid curve indicates this magnitude on the quarter circle of radius of $2\lambda_0$ where the detection points are located.  Moreover, we show (as blue triangle) the theoretical results for such scattered field amplitude as obtained from the analytical expression (Eq.~(\ref{Escat_Einc})) using Matlab \textsuperscript{\textregistered}, and the simulation results from the outputs of our proposed metadevice (shown as red circles).  We note the good agreement among all three approaches (for the phase information and the comparison, see the supporting information).  Fig.~(\ref{mainResult}) (c), shows the electric field distribution inside the inverse-designed metastructure and the relative mode amplitudes at the output ports (the results of which are shown by red circles in Fig.\ref{mainResult} (b)).  
\newline
Since our metadevice is designed based on Green's function formulation in Eq.~(\ref{Escat_Einc}), the incident field could be arbitrarily chosen. Therefore, the same metadevice is capable of computing the scattered fields for all sorts of incident radiation, without a need to be re-designed. For demonstrating this capability, as the second set of examples here, we displace the line source and evaluate the new scattered fields. Fig.~(\ref{variousResults_epsilon_Einc}) (a) \& (b) show the results for the line source displaced $0.2\lambda_0$ and $0.4\lambda_0$ downward relative to the center of the array, respectively. The panels next to the scattered field results show the electric field inside the inverse-designed metastructure and the mode amplitudes at the output ports.  The good agreement among the scattered field data confirms the ability of our metadevice to evaluate scattering fields for this scattering problem for an arbitrary incident radiation.  (For the phase information, see the supporting information.)

As mentioned earlier, our DCM stage is reconfigurable, and consequently it can be adjusted for different values of relative permittivities of the rods as long as the rods stay at their given locations.  This latter condition, when added to the fact that the detection points are also fixed, guarantees that the inverse-designed metastructure can be re-used without any need for redesigning.  This allows our metadevice to compute the scattered fields at the detection points for arbitrary relative permittivities of rods and arbitrary incident radiation.  This capability could be useful in optimization problems where one needs to evaluate the EM scattering for numerous different distributions of the relative permittivity of the object, for example in the ground-penetrating radar (GPR) and subsurface scenarios involving inverse-profiling of composite targets~\cite{zhang2018mimo,Insernia2011}. As the 3rd set of examples, for the same geometrical configuration of Fig.~(\ref{mainResult}) (a), we consider $\varepsilon_{r} = \{4,3,2,3,4\}$ as a new set of dielectric constants for the rods. Next, we modify some of the reconfigurable diagonal elements in the middle stage of our DCM architecture to implement the new matrix to be inverted in order to provide us with a new set of the polarization induced inside the objects according to their given relative permittivities. Fig.~(\ref{variousResults_epsilon_Einc}) (c) shows the scattered field results for the new set of relative permittivities of the objects being illuminated by a line source $\lambda_0 /2$ away from the center of the array. Fig.~(\ref{variousResults_epsilon_Einc}) (d) shows the corresponding results for anther set of relative permittivities of $\varepsilon_{r} = \{-4+j0.02,\ -2+j0.01,\ j0.01,\ -2+j0.01,\ -4+j0.02\}$. Again good agreement is noted among the scattered field magnitude results for both sets of relative permittivities, hence, verifying the capability of our metadevice to compute the scattered fields in forward-scattering problems with various relative permittivity profiles of the target scatterer. (for the phase information and the comparison, see the supporting information.)
\clearpage
As the 4th example, we demonstrate the capability of our metadevice for performing forward-scattering computations for through-the-wall imaging applications.  The through-the-wall radar imaging (TWRI) is a technique for making images of targets behind visually opaque dielectric walls using microwaves. The scattered fields from targets propagate back to the receiver for post-processing ~\cite{zhang_hoorfar_thajudeen_2018,amin_ahmad_2013,yemelyanov_engheta_hoorfar_mcvay_2009,baranoski_2008,zhang_hoorfar_2014}.  One of the challenges of the TWRI is the computation time (particularly when many iterations in the forward scattering are needed in such computation), which is a determining factor in the applicability of an imaging method for real-time applications~\cite{zhang_hoorfar_2013,zhang_hoorfar_thajudeen_2018}.  Here we show how an extension of our proposed metadevice can be utilized to do the forward-scattering computation for the TWRI. Since our device computes with waves (and thus may exhibit high wave speed), the required time for image reconstruction may be reduced considerably, especially if one is also interested in estimating the permittivity profiles of the targets behind or within the wall. As an aside, it is worth noting that in inverse profiling (e.g., GPR and subsurface applications), where one is concerned with finding both location and shape of an object as well as its material composition, the forward scattering problem becomes time consuming, even when the target location is fixed~\cite{imani_gollub_yurduseven_diebold_boyarsky_fromenteze_pulido-mancera_sleasman_smith_2020}. In such scenarios, our proposed meta-device may become even more useful in reducing computation time.

\begin{figure}[h]
    \centering
    \includegraphics[width = 1\textwidth]{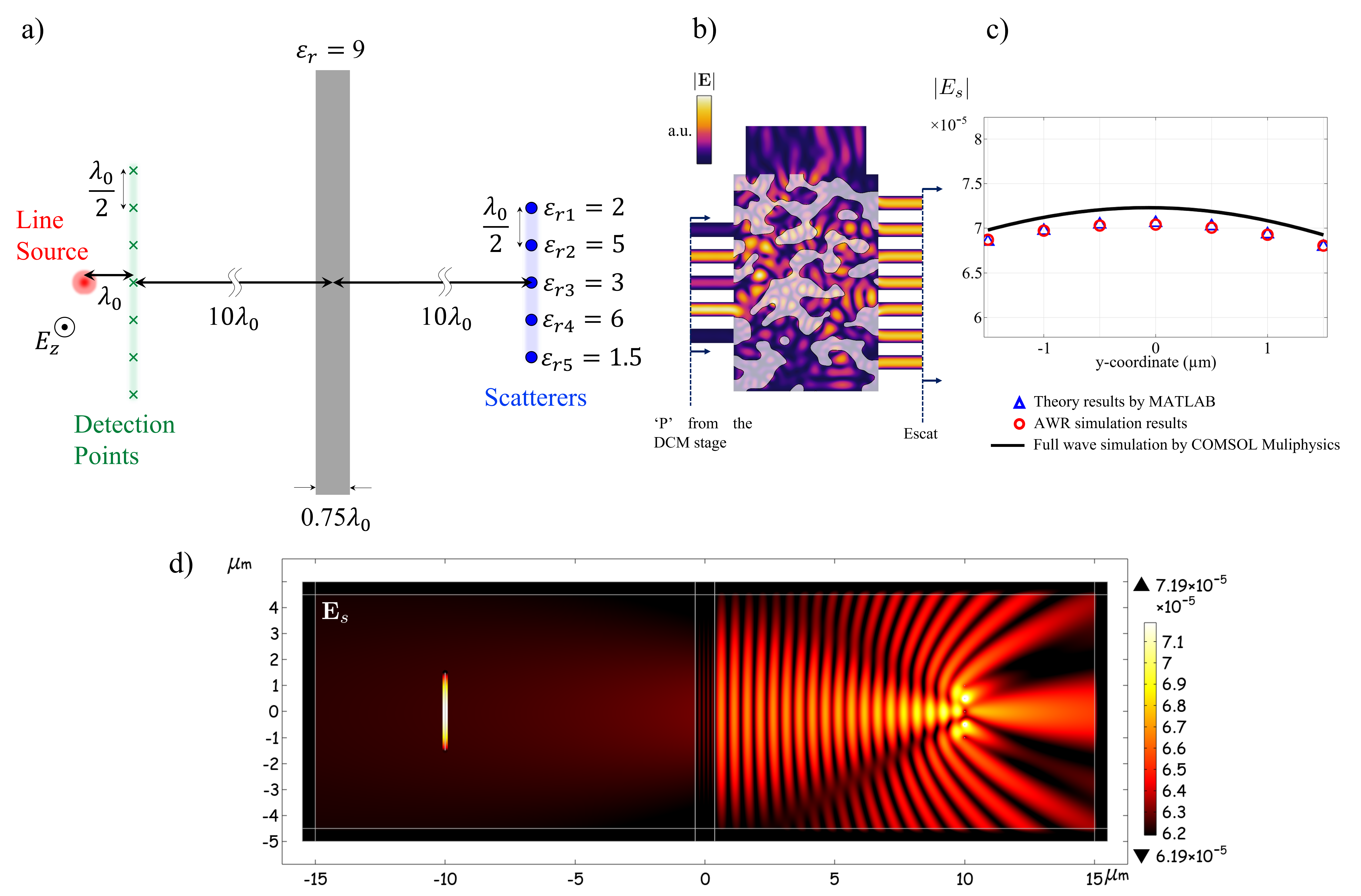}
    \caption{{\textbf{Through-the-wall-radar imaging (TWRI) application}}:  (a) The geometry of a through-the-wall forward scattering problem. 5 dielectric rods with spacing of $\lambda_0 /2$ and relative permittivities of $\varepsilon_r = \{2,5,3,6,1.5\}$ from top to bottom, respectively, are hidden behind a dielectric wall with $\varepsilon_r = 9$ and thickness $d = 0.75\lambda_0$. An infinitely line electric line source illuminates the scene and the scattered fields are detected at 7 designated detection points with spacing of $\lambda_0 /2$. The line source, detection points, and the scatterers are several wavelengths ($\sim 10\lambda_0$) away from the wall in order to use the far-field Green's function approximation in the presence of the wall. (b) the inverse-designed metastructure for emulating such Green's function, $\vec{G}^\t{w}$, for this problem (a). The input ports are excited with the complex amplitudes proportional to the induced diepole polarizations computed from the DCM stage. (c) The scattered field data at the detection points are evaluated with full-wave simulations in COMSOL Multiphysics\textsuperscript{\textregistered} (black solid curve), in Eq.~(\ref{Wall_Escat_Einc}) in Matlab\textsuperscript{\textregistered} (blue triangles), and in the outputs of the metastructure simulated in AWR Microwave Office\textsuperscript{\textregistered} (red circles) (d) The magnitude of the scattered electric field distribution simulated for the EM model shown in panel(a).}
    \label{TWRI_Results}
\end{figure}

A typical EM model of through-the-wall forward scattering for a single-layer dielectric wall is shown in Fig.~(\ref{TWRI_Results}) (a). A 2D infinitely long line source with an out-of-plane electric field illuminates the scene and the detection points are positioned close to the source in a linear configuration, for receiving the scattered fields from the 2D target scatterers behind the wall. In order to use the far-field approximation of the wall's Green function the line source, scatterers, and the detection points are located several wavelengths away from the wall. The far-field Green function for a single-layer dielectric wall is as follows:

%--------------------------------------------------
\begin{equation}\label{Wall_GreensFunc}
     \vec{G}^\t{w}(\vec{r}-\vec{r}_s) \approx T_W \vec{G}^\t{pr}(\vec{r}-\vec{r}_s)=-jT_W\frac{k_0^2}{4\pi\varepsilon_0}H^{(2)}_0(k_0|\vec{r}-\vec{r}_s|)
\end{equation}
%--------------------------------------------------

where $\vec{G}^\t{w}$ and $\vec{G}^\t{pr}$ are the 2D Green function in the presence of the wall and the 2D Green function in free-space, respectively. $T_\t{w}$ is the wall's transmission coefficient for a normal incident plane wave and $\vec{r}_s$ is the position of the line-source. 
The transmitted fields through the wall illuminate the objects behind it and the scattered fields propagate back through the wall to reach the detection points. Therefore, using the notation we used earlier, we may write the following formulation for calculating the scattered fields at the designated detection points: 

%--------------------------------------------------
\begin{equation}\label{Wall_Escat_Einc}
    e_\t{sca}=T_\t{w} G^\t{pr}~p= T_\t{w} G^\t{pr}\left(\t{diag}(\alpha^{-1})-G\right)^{-1} T_\t{w} e_\t{inc}
\end{equation}
%--------------------------------------------------

Compared to the Eq.~(\ref{Escat_Einc}) for a scattering problem in a free-space background, the only change we need to make for the present case with the wall is to include the transmission coefficient twice, one for the incident wave going through the wall to reach the scatterers and the other is for the scattered fields from the objects to reach the detection points. We note that the Green function in Eq.~(\ref{Wall_GreensFunc})-(\ref{Wall_Escat_Einc}) may be appropriately modified to redesign and tailor the metadevice for other forward scattering problems of interest such as those in intra-wall imaging~\cite{zhang_hoorfar_thajudeen_2018} on in GPR subsurface profiling~\cite{hajebi_hoorfar_2021}.

In Fig.~(\ref{TWRI_Results}) (a), we consider a case of through-the-wall forward scattering. The thickness and relative permittivity of the wall are assumed to be $0.75 \lambda_0$ and $\varepsilon_r = 9$, respectively. Five 2D dielectric rods with diameter of $d = \lambda_0/50$ are positioned as a linear array $10 \lambda_0$ away from the wall. The distance between adjacent cylinders is $\lambda_0 /2$ and the relative permittivities of the rods are $\varepsilon_{r} = \{2,\ 5,\ 3,\ 6,\ 1.5\}$. Seven detection points with spacing of $\lambda_0/2$ are on the other side of the wall close to the source. The source is positioned $\lambda_0$ away from the detection points. Fig.~(\ref{TWRI_Results}) (d) depicts the results of the full-wave simulation for the scattered field distribution using COMSOL Multiphysics \textsuperscript{\textregistered} ~\cite{COMSOL}. Fig.~(\ref{TWRI_Results}) (b) presents the inverse-designed metasrtcuture for emulating the far-field Green function with the wall present in the configuration shown in Fig.~(\ref{TWRI_Results}) (a). Fig.~(\ref{TWRI_Results}) (c) shows the magnitude of scattered fields at the detection points using the COMSOL Multiphysics \textsuperscript{\textregistered} full-wave simulation (black solid curve), the numerical results of Eq.~(\ref{Wall_Escat_Einc}) implemented in Matlab\textsuperscript{\textregistered} (blue triangles), and the simulation results for our metadevice (red circles). Close agreement is observed, verifying the capability of our proposed metadevice for performing forward-scattering computations for through-the-wall imaging. (For phase information and the comparison, see the supplementary information.)
\clearpage
\section{Summary and Conclusion}
In this work, we proposed a wave-based metadevice for computing electromagnetic forward scattering problems. The scattering object on which the integral equation is defined is discretized into small cells and the induced dipole polarizations inside the cells can be computed by inverting a matrix. We proposed a mesh of tunable couplers based on the DCM architecture for computing the matrix and unity feedback on top of it for inverting it. Then the computed polarizations are fed into an inverse-designed metastructure for determining the scattered fields at the designated detection points using the propagator Green's function. The proposed metadevice is capable of computing the scattered fields for various scattering problems with different relative permittivity profiles of the object and various different incident excitations. Although our proposed platform is customized for solving electromagnetic scattering problems, we can use the general concept behind this architecture to design other wave-based metastructure for solving other physical problems. One of the interesting potentials of this wave-based computing platform is the possibility for enhanced computation speed offered by the high speed of electromagnetic waves. Therefore, in applications in which numerical computation of a specific physical problem needs to be performed many times, our architecture may decrease the wait time for computations significantly.

\section*{Acknowledgements}
This work is supported in part by the US Air Force Office of Scientific Research (AFOSR) Multidisciplinary University Research Initiative (MURI) grant number FA9550-17-1-0002, and in part by the US National Science
Foundation (NSF) MRSEC program under award No. DMR-1720530.

\section*{Notes}
The authors declare no competing financial interests.  N.E. is a strategic scientific advisor/consultant to Meta Materials, Inc. The numerical data, codes, and other data that support the findings of this study are available from the authors upon reasonable request.

\newpage
\bibliographystyle{ieeetr}
\bibliography{bibliography.bib}

\end{document}

\title{Supporting Information \\ Inverse-designed Metastructures together with Reconfigurable Couplers to Compute Forward Scattering}
\author{Vahid Nikkhah}\thanks{These authors contributed equally}
\affiliation{University of Pennsylvania, Department of Electrical and Systems Engineering, Philadelphia, PA 19104 U.S.A.}
\thanks{These authors contributed equally}
\author{Dimitrios Tzarouchis}\thanks{These authors contributed equally} 
\affiliation{University of Pennsylvania, Department of Electrical and Systems Engineering, Philadelphia, PA 19104 U.S.A.}
\thanks{These authors contributed equally}
\author{Ahmad Hoorfar} 
\affiliation{Villanova University, Department of Electrical and Computer Engineering, Villanova, PA 19085 U.S.A.}
\author{Nader Engheta}
\affiliation{University of Pennsylvania, Department of Electrical and Systems Engineering, Philadelphia, PA 19104 U.S.A.}
% \date{\today}
\maketitle

\section{Normalizing the DDA formulation}
Our proposed wave-based metastructure is designed to emulate the discrete dipole approximation (DDA) formulation of a forward scattering problem as follows  ~\cite{Purcell1973,Draine1994}:
\begin{equation}\label{DDA}\tag{S1}
	e_\text{scat}=G^\text{pr}\left(\text{diag}(\alpha^{-1})-G\right)^{-1}e_\text{inc}
\end{equation}
where $G_{ik}^\text{pr}=-j\frac{k_0^2}{4\pi \varepsilon_0}H^{(2)}_0(k_0|\vec{r}_i-\vec{r}'_k|)$, $G_{ik}=-j\frac{k_0^2}{4\pi \varepsilon_0}H^{(2)}_0(k_0|\vec{r}'_i-\vec{r}'_k|),\ \ G_{ii}=0$ are the Green function elements and $\vec{r}_i$ , $\vec{r}'_i$ are the position vectors of detection point $i$ and target object $i$, respectively. On the one hand, the modulus of the coefficient $s = -j\frac{k_0^2}{4\pi \varepsilon_0}$ in front of the Hankel functions $H_0^{(2)}(.)$ can attain very large values e.g., $|s| \simeq 3.55\times 10^{23}$ for $\lambda_0 = 1 \ \mu m$. On the other hand the polarizabilities $\alpha = A_\text{rod} \varepsilon_0 (\varepsilon_r - 1) = \frac{\pi d^2}{4}\varepsilon_0 (\varepsilon_r - 1) $ may have very small values, e.g., $\alpha \simeq 5.56\times10^{-27}$ for $\lambda_0 = 1 \ \mu m$, $d = \lambda_0 /50$ and $\varepsilon_r = 2$. Since the values of the quantities that are involved in Eq.~(\ref{DDA}) are distributed over vast orders of magnitude, one cannot implement Eq.~(\ref{DDA}) in its current form. Therefore, we rewrite Eq.~(\ref{DDA}) such that the physical quantities of interest that are multiplied by the incident radiation become normalized and dimensionless. We start with the Green function matrices and define the normalized elements as:
\begin{equation}\label{DDA_normG}\tag{S2}
	\begin{split}
		& \widetilde{G}_{ik}^\text{pr} = H^{(2)}_0(k_0|\vec{r}_i-\vec{r}'_k|) \\
        & \widetilde{G}_{ik}= H^{(2)}_0(k_0|\vec{r}'_i-\vec{r}'_k|) \\
        & \Rightarrow G_{ik}^\text{pr} = s\widetilde{G}_{ik}^\text{pr} ,  G_{ik} = s\widetilde{G}_{ik}\\ 
        & e_\text{scat}=s\widetilde{G}^\text{pr}\left(\text{diag}(\alpha^{-1})-s\widetilde{G}\right)^{-1}e_\text{inc} \\
        & \Rightarrow e_\text{scat}=\widetilde{G}^\text{pr}\left(\frac{1}{s}\text{diag}(\alpha^{-1})-\widetilde{G}\right)^{-1}e_\text{inc}
	\end{split}
\end{equation}
where $\widetilde{G}_{ik}^\text{pr}$ and $\widetilde{G}_{ik}$ are the normalized dimensionless Green function elements and $s = -j\frac{k_0^2}{4\pi \varepsilon_0}$ is the scalar quantity in front of the normalized values that gives the actual Green's function elements. Canceling out the scalar $s$ leaves behind the normalized elements of Green functions as shown in the last equation. Now, we proceed further to normalize the scalar quantity itself and the polarizabilities by normalizing the product $\frac{1}{s \alpha}$ as shown in the following:
\begin{equation}\label{DDA_norm}\tag{S3}
	\begin{split}
		& \frac{1}{s\alpha}  = \frac{1}{-j \frac{k_0^2}{4\pi\varepsilon_0} \frac{\pi d^2}{4} \varepsilon_0(\varepsilon_r - 1)} = \frac{1}{-j \pi \times \frac{\pi}{4} (\frac{d}{\lambda_0})^2 (\varepsilon_r -1)} = \frac{1}{\widetilde{s}\widetilde{\alpha}} \\
        & \widetilde{s} = -j\pi, \ \ \widetilde{\alpha} = \frac{\pi}{4} (\frac{d}{\lambda_0})^2 (\varepsilon_r -1) \\
        & \Rightarrow e_\text{scat}=\widetilde{G}^\text{pr}\left(\frac{1}{\widetilde{s}}\text{diag}(\widetilde{\alpha}^{-1})-\widetilde{G}\right)^{-1}e_\text{inc} = \widetilde{G}^\text{pr} \widetilde{p}
	\end{split}
\end{equation}
As can be seen the scalar $s$ and polarizabilities $\alpha$ are normalized simultaneously leaving behind the normalzied version of the DDA as written in the last equation. This normalized version of the DDA is implemented by our wave-based metastructure.
\newpage
\section{Phase of the scattered field at the detection points}
All the physical quantities involved in the DDA formulation of Eq.~(\ref{DDA}) such as incident radiation vector and the Green function and polarizablity matrices  have phase information, hence complex-valued quantities. The metastructure is a wave-based coherent system allowing the complex-valued quantities to be encoded on the magnitude and phase of the electromagnetic waves. In the main text, it is shown that the magnitude of the scattered fields at the detection points calculated by the metastructure matches well with the magnitude of the scattered fields computed from full-wave simulations in COMSOL Multyphysics\textsuperscript{\textregistered} and the direct numerical inversion of the matrices in the DDA approximation. In this section, we present the phase of the scattered fields,$\phi_\text{scat} = \tan^{-1} \left \{ \Im(E_\text{scat})/\Re(E_\text{scat})\right\}$, at the detection points computed from the outputs of the metastructure, and compared them with the numerical values from full-wave simulations and DDA formulation. Fig.~\ref{phase_re_im_config1}-\ref{phase_re_im_config5} illustrate the phase data for the 5 scattering problems analyzed in the main text. Panel (a) in each figure shows the phase of the scattered field at the detection points in a linear plot and panels (b) and (c) depict the real and imaginary parts of the scattered field. Black curve represents the values from full-wave simulations and the blue triangles and the red circles are the values computed from DDA formulation and the outputs of the metastructure, respectively. The error between the phase values from the outputs of the metastructure and the other simulation values does not exceed $\delta \phi_\text{scat} = 5^\circ$.
\renewcommand{\thefigure}{S1}
\begin{figure}[h]
    \centering
    \includegraphics[width = \textwidth]{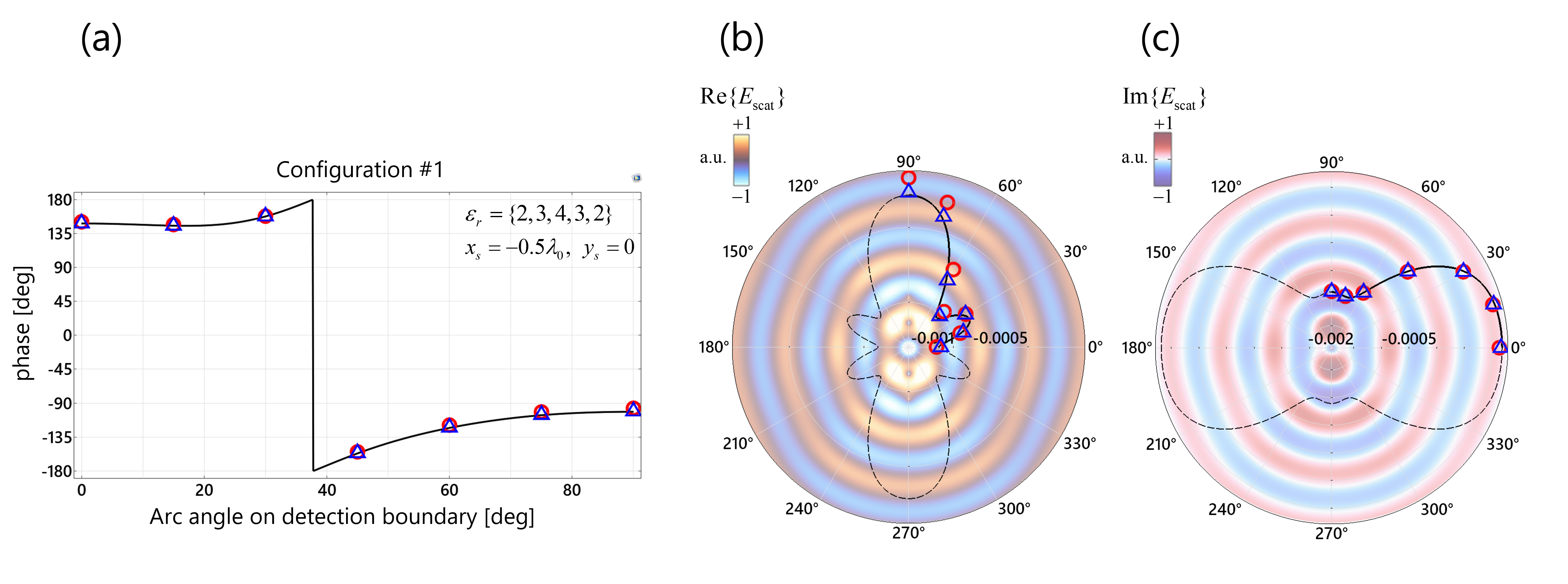}
    \caption{(a) Phase data of the scattered field for example 1 (see the main text for the details). The black curve represents the phase data from the full-wave simulation in COMSOL Multiphysics\textsuperscript{\textregistered}. The blue triangles and the red circles are the phase data computed from the DDA formulation and the simulation results of the outputs of the metastructure, respectively. The 2D scattered field distributions in panels (b) and (c) are the real and imaginary parts computed from full-wave simulations, respectively.}
    \label{phase_re_im_config1}
\end{figure}
\renewcommand{\thefigure}{S2}
\begin{figure}[h]
    \centering
    \includegraphics[width = \textwidth]{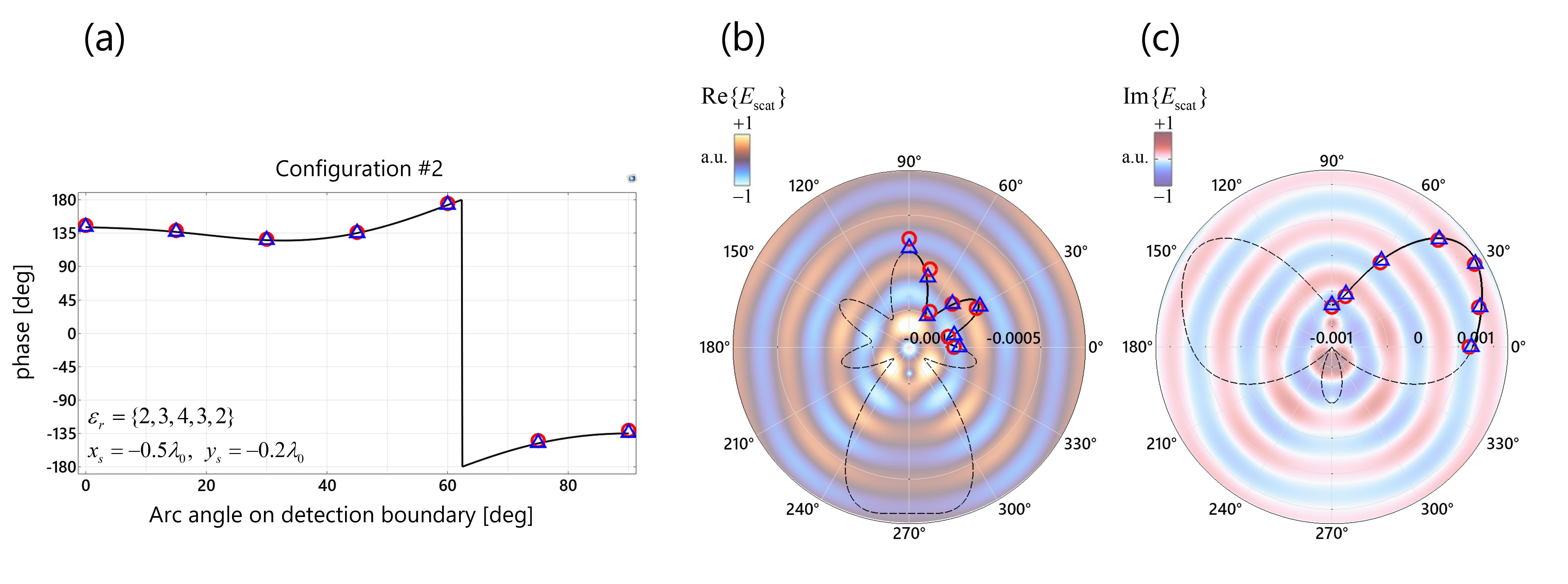}
    \caption{Phase data of the scattered field for example 2}
    \label{phase_re_im_config2}
\end{figure}
\renewcommand{\thefigure}{S3}
\begin{figure}[h]
    \centering
    \includegraphics[width = \textwidth]{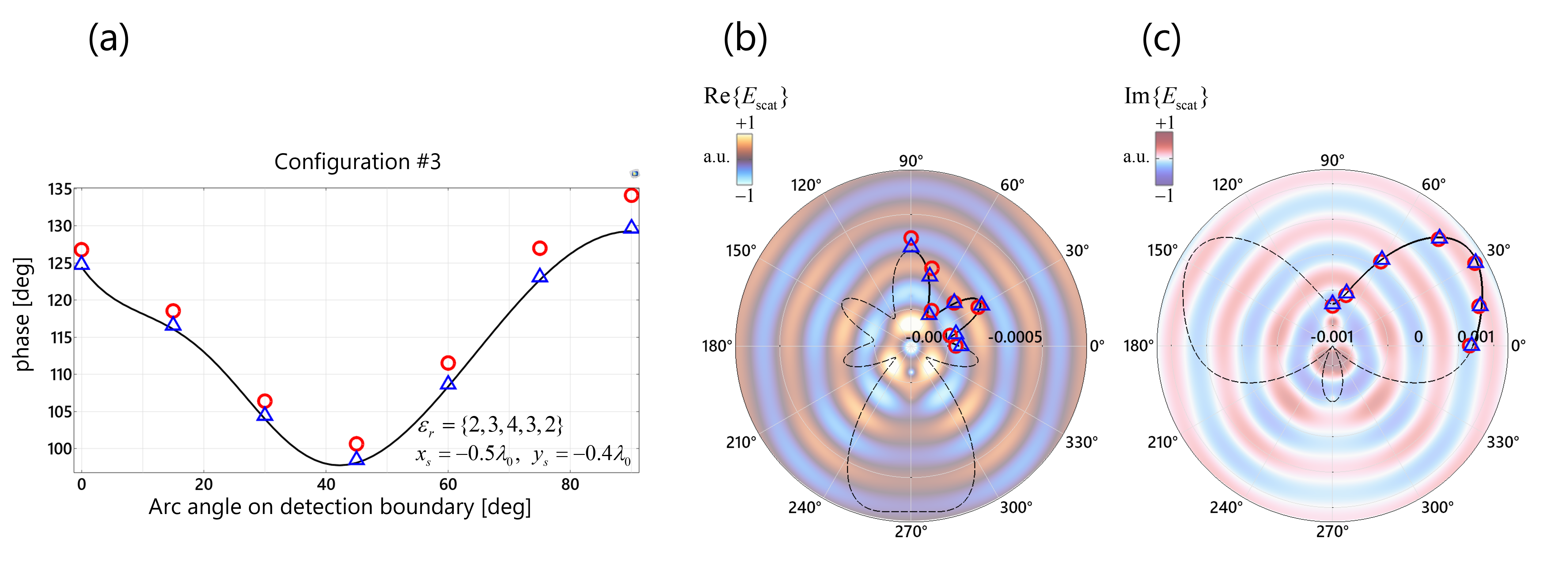}
    \caption{Phase data of the scattered field for example 3}
    \label{phase_re_im_config3}
\end{figure}
\renewcommand{\thefigure}{S4}
\begin{figure}[h]
    \centering
    \includegraphics[width = \textwidth]{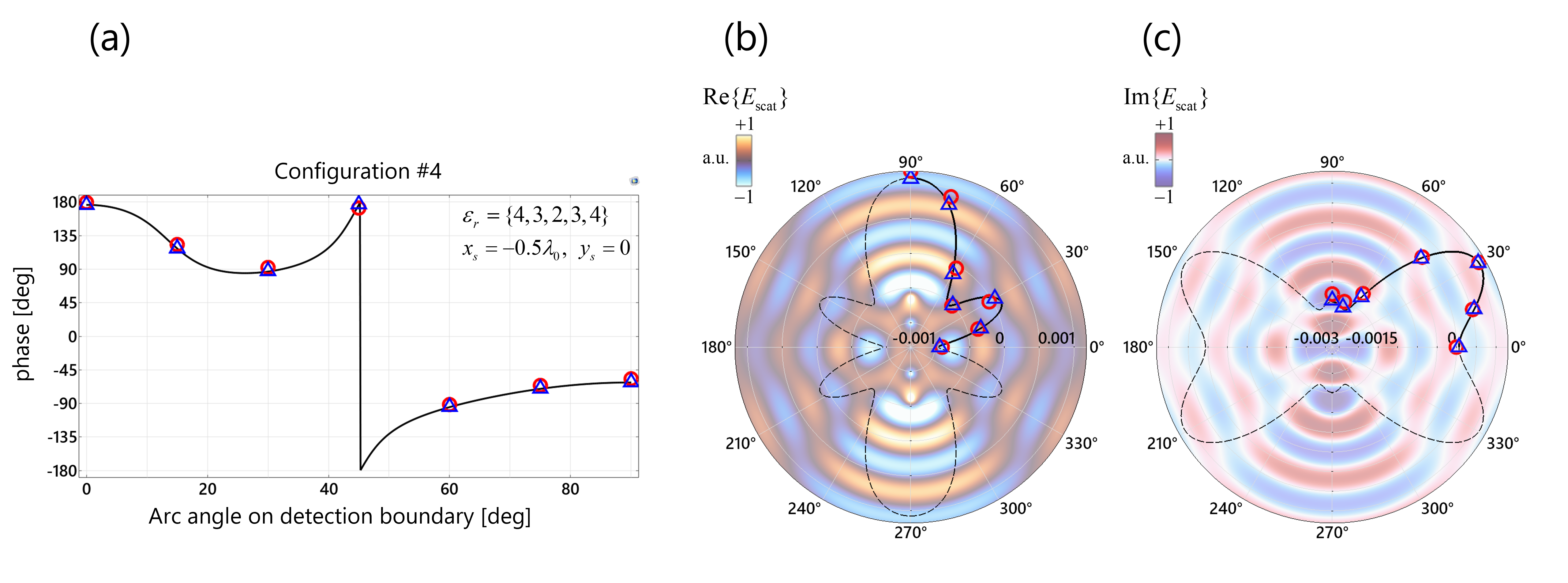}
    \caption{Phase data of the scattered field for example 4}
    \label{phase_re_im_config4}
\end{figure}
\renewcommand{\thefigure}{S5}
\begin{figure}[h!]
    \centering
    \includegraphics[width = \textwidth]{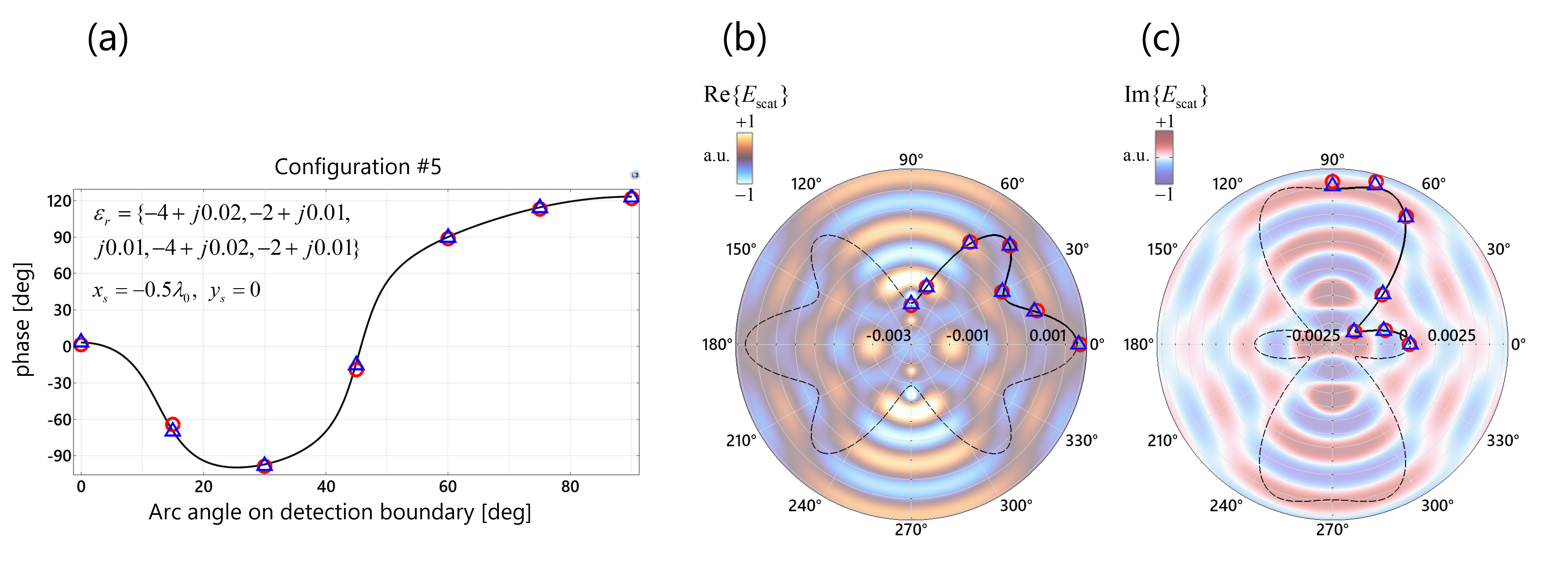}
    \caption{Phase data of the scattered field for example 5}
    \label{phase_re_im_config5}
\end{figure}
\clearpage
\newpage
\section{details of the Inverse design}
For implementing $\widetilde{G}^\text{pr}$, we utilize a density-based topology optimization technique ~\cite{molesky_2018,bendsoe_sigmund_2019,sigmund_chigrin_2009} using a gradient-descent approach implemented by adjoint sensitivity analysis in COMSOL. In this technique, a normalized continuous density function $0 \le \rho(\vec{r}) \le 1$ is defined over the design region, e.g. the hallowed metallic box in the microwave network, and an objective scalar defining the figure of merit (FOM) of the performance is optimized relative to the density function. The density function in this case is mapped to the relative permittivity distribution inside the design region. For removing small feature size domains and binarizing the density function for having a physically-realizable material distribution, spatial filtering and a projection function are applied on the density function, $\rho(\vec{r})$, as following:
\begin{equation}\label{desnity_filter_projection}\tag{S4}
    \begin{split}
        \rho_\text{f} (\vec{r}) = \rho(\vec{r}) + R_\text{min} \nabla^2 \rho_\text{f}(\vec{r}) \\
    \rho_\text{p} (\vec{r}) = \frac{1}{1+\text{exp}(-s(\rho_\text{f}(\vec{r})-0.5))}
    \end{split}
\end{equation}
where $\rho_\text{f} (\vec{r})$ and $\rho_\text{p} (\vec{r})$ are the filtered and projected density functions, respectively. The first equation is a Helmholtz-type spatial filter, which filters the feature size domains smaller than $R_\text{min}$. The steepness factor $s$ in the projection function determines the sharpness of the transition of $\rho_\text{f} (\vec{r})$ from 0 to 1 around the middle value $\rho_\text{f} (\vec{r}) =0.5$. The larger the $s$ parameter, the more binarized the material distribution. Finally, the relative permittivity distribution inside the design region is interpolated by a mapping function such as:
\begin{equation}\label{epsilon_density}\tag{S5}
   \varepsilon_r(\vec{r}) = \varepsilon_{r,\text{air}} + (\varepsilon_{r,\text{poly}} - \varepsilon_{r,\text{air}}) \rho_\text{p}(\vec{r})
\end{equation}
For instance, the above mapping function interpolates the density between air $\varepsilon_{r,\text{air}} = 1$ where $\rho_\text{p}(\vec{r}) = 0$ and polystyrene $\varepsilon_{r,\text{poly}} = 2.53$ where $\rho_\text{p}(\vec{r}) = 1$. In setting up the optimization problem, the wave equation for the electric field is used as the constraint equation for the electric field inside the simulation domain, and the electric field at the ports boundaries is used to define the objective scalar based on the desired performance. For time-harmonic electromagnetic fields the wave equation may be written as following:
\begin{equation}\label{Maxwells_Eq}\tag{S6}
   \nabla \times \mu^{-1}_r (\nabla \times \vec{E}_n) -\omega^2_0 \mu_0 \varepsilon_0 \varepsilon_r(\vec{r})\vec{E}_n = -i \omega_0 \vec{J}_n
\end{equation}
where $\vec{E}_n$ is the electric field generated by $\vec{J}_n$, the source term in Eq.(\ref{Maxwells_Eq}). $\vec{J}_n$ is the equivalent current density of $\text{TE}^{10}$ mode at the input port $n$. Each input port is separately excited and the solution to Eq.(\ref{Maxwells_Eq}) is then used to define the objective scalar, which is written as following:
 \begin{equation}\label{Objective_Scalar}\tag{S7}
    \mathcal{S} = \sum_{n=1}^{5} \left( \sum_{m=1}^{7} \iint_{\partial V} |\vec{E}_{n} (\vec{r}_m)-T_{mn} \vec{E}^{10}(\vec{r}_m)|^2 d^2 r_m + \sum_{k=1}^{5} \iint_{\partial V} |\vec{E}_{n} (\vec{r}_k) -\delta_{nk}\vec{E}^{10}(\vec{r}_k) |^2 d^2 r_k \right) 
\end{equation}
In the first part of the objective scalar $\mathcal{S}$, $\vec{E}_n(\vec{r}_m)$ is the electric field on the cross section of the output port $m$ when the input port $n$ is excited. $\vec{E}^{10}(\vec{r}_m)$ is the normalized electric field of the fundamental $\text{TE}^{10}$ mode at the cross section of the output port $m$ and $T_{mn}$ is the target transmission coefficient from the input port $n$ to the output port $m$. In the second part, $\vec{E}_n(\vec{r_k})$ is the electric field at the input port $k$ when the input port $n$ is excited and similarly $\vec{E}^{10}(\vec{r}_k)$ is the electric field of the $\text{TE}^{10}$ mode on the input port $k$. $\delta_{nk}$ is the Kronecker delta symbol.
The objective scalar is defined based on the "distance" (in the complex domain) between the complex-valued electric field from Eq.(\ref{Maxwells_Eq}) for the current material distribution inside the design region and the target complex-valued electric field on the ports cross section based on the desired functionality. In other words, the objective scalar is the magnitude of the complex-valued error field between the current and target electric fields on the ports. The first part of the objective scalar $\mathcal{S}$ tries to set the complex mode amplitudes of the $\text{TE}^{10}$ mode at the output ports according to the elements of the $\gamma \widetilde{G}^\text{pr}$ matrix while each input port is excited with an unit amplitude $1e^{j0}$. The purpose of the scalar $\gamma$ in front of the $\gamma \widetilde{G}^\text{pr}$ matrix will be described shortly. The second part of the objective scalar is designed to minimize the reflection towards input port $k$ and the coupling from input port $k$ to other input ports $n\neq k$.

The proposed wave-based metastructure is a passive device capable of implementing transmission matrcies that results in output power smaller than (or at best equal to) the input power. To determine whether a given matrix $\widetilde{G}^\text{pr} \in \mathbb{C}^{n\times m}$ can be implemented or not, we use singular value decomposition (SVD) of $\widetilde{G}^\text{pr}$ matrix and calculate its singular values, $\sigma_i(\widetilde{G}^\text{pr})\ \ i=1,2,..,\min \{m,n\}$. If the largest singular value, $\sigma_\text{max} (\widetilde{G}^\text{pr})$, is smaller than 1 then $\widetilde{G}^\text{pr}$ represents a passive transmission, which is realizable by a passive microwave/optical network. Otherwise, if $\sigma_\text{max} > 1$ then $\widetilde{G}^\text{pr}$ needs to be scaled down by some scalar $\gamma<1$. Scaling down the matrix results in scaling down all the singular values by $\gamma$ i.e., $\sigma_i(\gamma \widetilde{G}^\text{pr}) = \gamma \sigma_i(\widetilde{G}^\text{pr})$. Therefore, we can calculate the scalar $\gamma$ such that the maximum singular value of the scaled matrix becomes smaller than unity, hence becoming realizable by a passive network:
\begin{equation}\tag{S8}
	\begin{split}
		& \sigma_\text{max} (\gamma \widetilde{G}^\text{pr}) = \gamma \sigma_\text{max} (\widetilde{G}^\text{pr}) < 1 \\
        & \Rightarrow \gamma < \min \left(1, \frac{1}{\sigma_\text{max} (\widetilde{G}^\text{pr})} \right)
	\end{split}
\end{equation}
For instance, from the configuration of the scatterers and the detection points in the first forward scattering example analyzed in the main text, $\widetilde{G}^\text{pr}$ is calculated as below:
\begin{equation}\label{G_pr}\tag{S9}
\widetilde{G}^\text{pr} =
    \begin{pmatrix}
    0.213e^{-i0.69}  &  0.222e^{i0.41}   &  0.225e^{i0.80}  &  0.222e^{i0.41}   &  0.213e^{-i0.69} \\
    0.203e^{-i2.08}  &  0.215e^{-i0.36}  &  0.225e^{i0.80}  &  0.229e^{i1.22}   &  0.225e^{i0.85} \\
    0.196e^{i3.02}   &  0.210e^{-i1.04}  &  0.225e^{i0.80}  &  0.237e^{i2.04}   &  0.242e^{i2.48} \\
    0.190e^{i2.06}   &  0.206e^{-i1.59}  &  0.225e^{i0.80}  &  0.245e^{i2.78}   &  0.262e^{-i2.18} \\
    0.187e^{i1.36}   &  0.203e^{-i2.01}  &  0.225e^{i0.80}  &  0.252e^{-i2.89}  &  0.286e^{-i0.70} \\
    0.184e^{i0.94}   &  0.202e^{-i2.26}  &  0.225e^{i0.80}  &  0.258e^{-i2.48}  &  0.308e^{i0.39} \\
    0.184e^{i0.79}   &  0.201e^{-i2.35}  &  0.225e^{i0.80}  &  0.260e^{-i2.34}  &  0.318e^{i0.81}
    \end{pmatrix}
\end{equation}
The singular values of the above matrix are $\sigma = [0.92,0.69,0.64,0.33,0.04]$, which already represents a passive transmission since $\sigma_\text{max}(\widetilde{G}^\text{pr}) = 0.92 < 1$. Therefore, we can directly implement $\widetilde{G}^\text{pr}$ without scaling the matrix.

\clearpage
\section{Details of the System Simulation }

Once the required metastructure is inverse-designed to give the required $G^\t{pr}$ matrix, the resulting S-parameters are inserted as a system block into the AWR Microwave Office\textsuperscript{\textregistered} as can be seen in the figure. Overall, the entire system consists of the following sub parts Fig.~\ref{AWR Schem} (the system is similar to the one introduced in~\cite{Tzarouchis2021})
\begin{itemize}
    \item the input/feedback couplers (Fig.~\ref{AWR Schem} (a) and (b) and Fig.~\ref{coupler}). This particular stage implements the following matrix
    \begin{equation}\tag{S10}
 S=\left[
 \begin{array}{cccc}
   0 & -\alpha_i & 0 & \beta_i \\
   -\alpha_i & 0 & \beta_i & 0 \\
   0 & \beta_i & 0 & \alpha_i \\
   \beta_i & 0 & \alpha_i & 0
 \end{array}\right]
 \end{equation}
 where $\alpha_i$ is the transmission coefficient and $\beta_i$ is the coupling between the ports~\cite{Pozar}, while $i=1,2$ indicating the two different couplers. In all examples the coupled power is $\beta_i^2=-30\t{dB}~(0.1\%)$. 
 
    \item the ingress, middle, and egress stage ((Fig.~\ref{AWR Schem} (c), (d), and (e))) - each stage is implemented with an MZI module (Fig.~\ref{MZI}). 
    \item The inverse-designed metastructure (Fig.~\ref{AWR Schem} (f) and Fig.~\ref{swiss}) 
\end{itemize}

\renewcommand{\thefigure}{S6}
\begin{figure}[h]
    \centering
    \includegraphics[width = 0.85\textwidth]{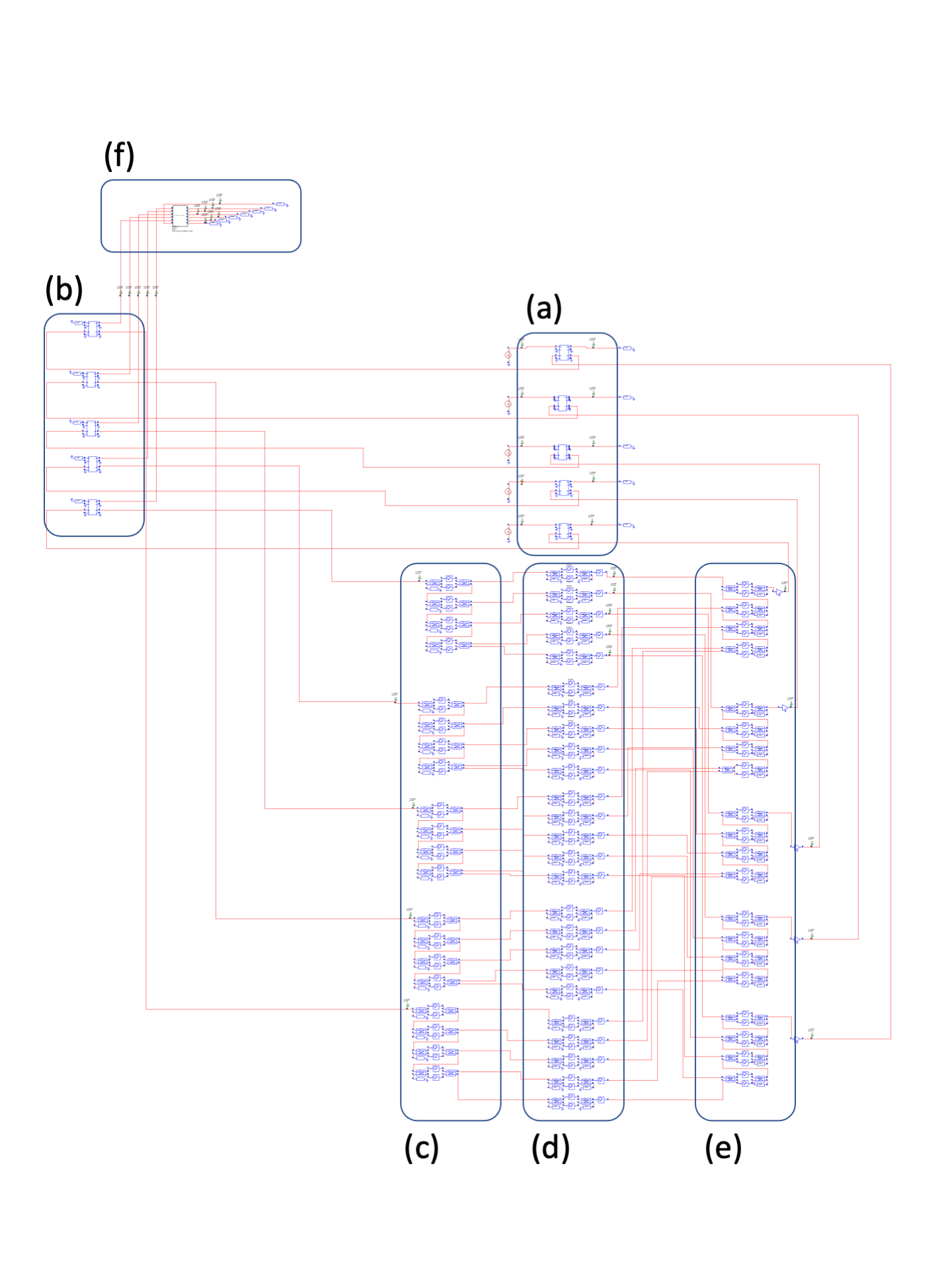}
    \caption{The system schematic in AWR Microwave Office\textsuperscript{\textregistered}. A vector of 5 inputs crosses (a) the first feedback coupler, continues at the second coupler (b), enters the (c) ingress, (d) middle and (e) egress stage and recombines at the second coupler (b). Finally the 5 inputs cross the inverse-designed metastucture, where 7 outputs deliver the required results}
    \label{AWR Schem}
\end{figure}

\renewcommand{\thefigure}{S7}
\begin{figure}[h]
    \centering
    \includegraphics[width = 0.85\textwidth]{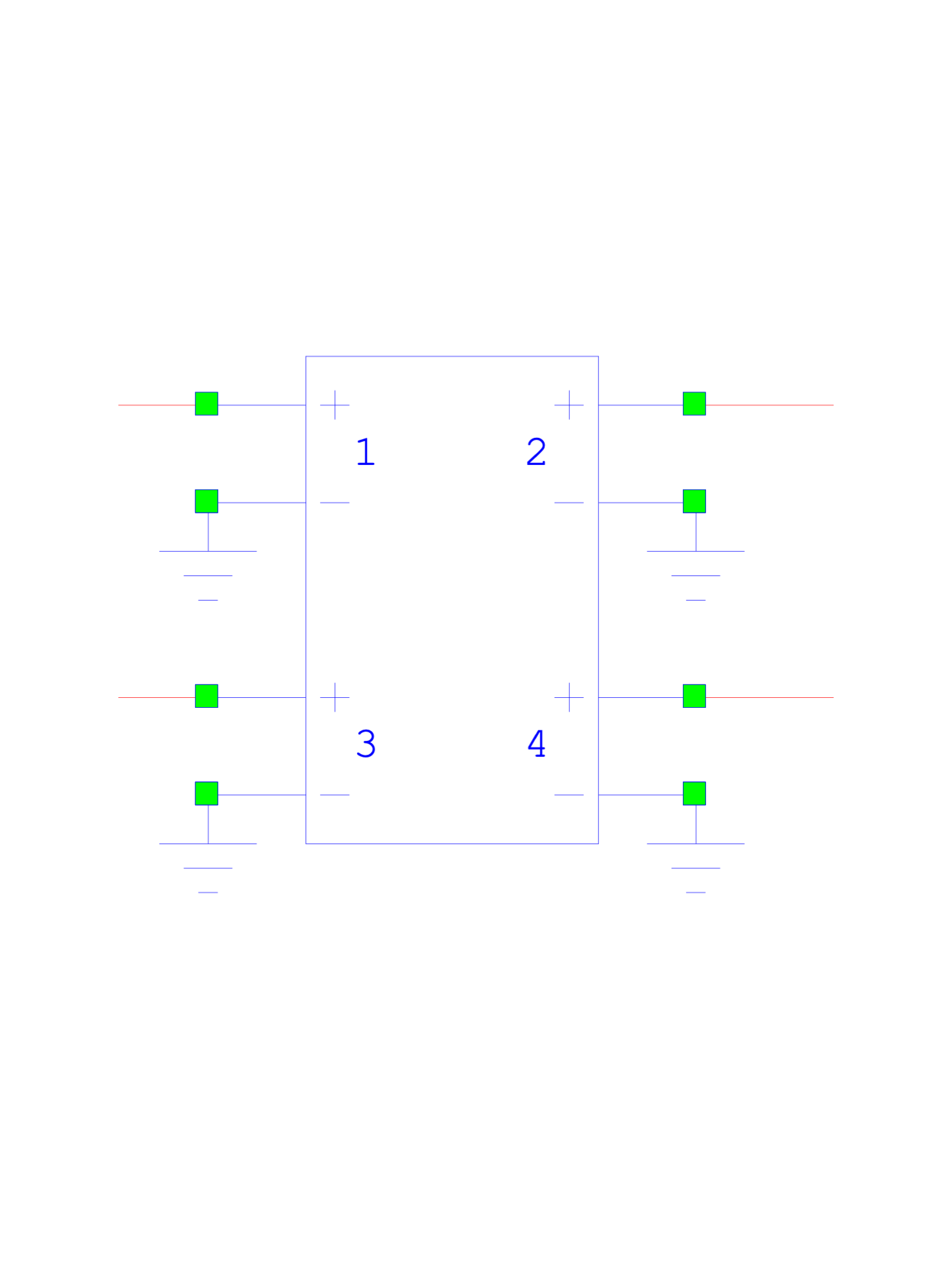}
    \caption{The system schematic of the utilized couplers, where 1 and 3 are the inputs while 2 and 4 the outputs, implemented in AWR Microwave Office\textsuperscript{\textregistered}, see also~\cite{Tzarouchis2021}.}
    \label{coupler}
\end{figure}

\renewcommand{\thefigure}{S8}
\begin{figure}[h]
    \centering
    \includegraphics[width = 0.85\textwidth]{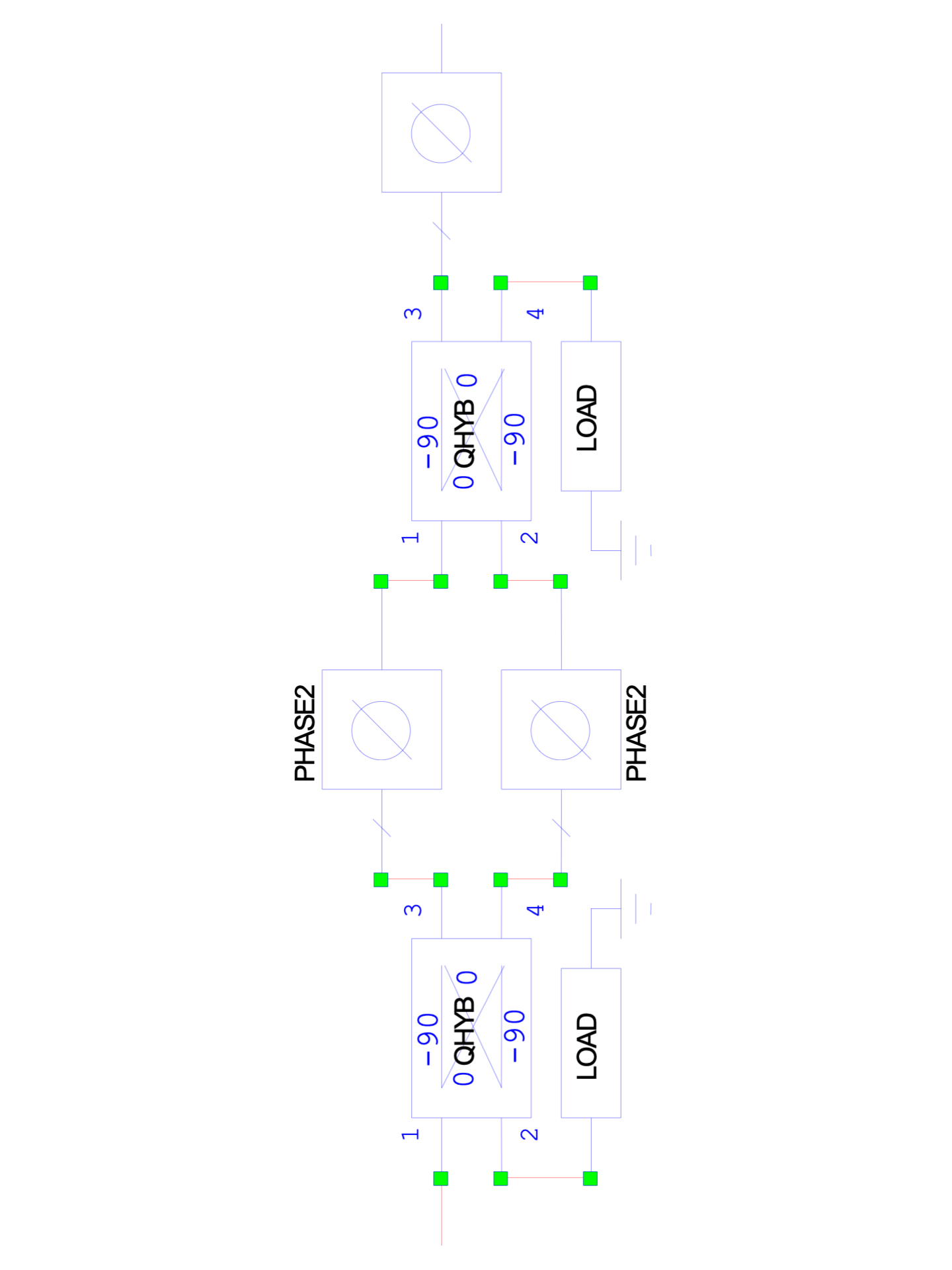}
    \caption{The system schematic of the MZI module, implemented in AWR Microwave Office\textsuperscript{\textregistered}, see also~\cite{Tzarouchis2021}.}
    \label{MZI}
\end{figure}

\renewcommand{\thefigure}{S9}
\begin{figure}[h]
    \centering
    \includegraphics[width = 0.85\textwidth]{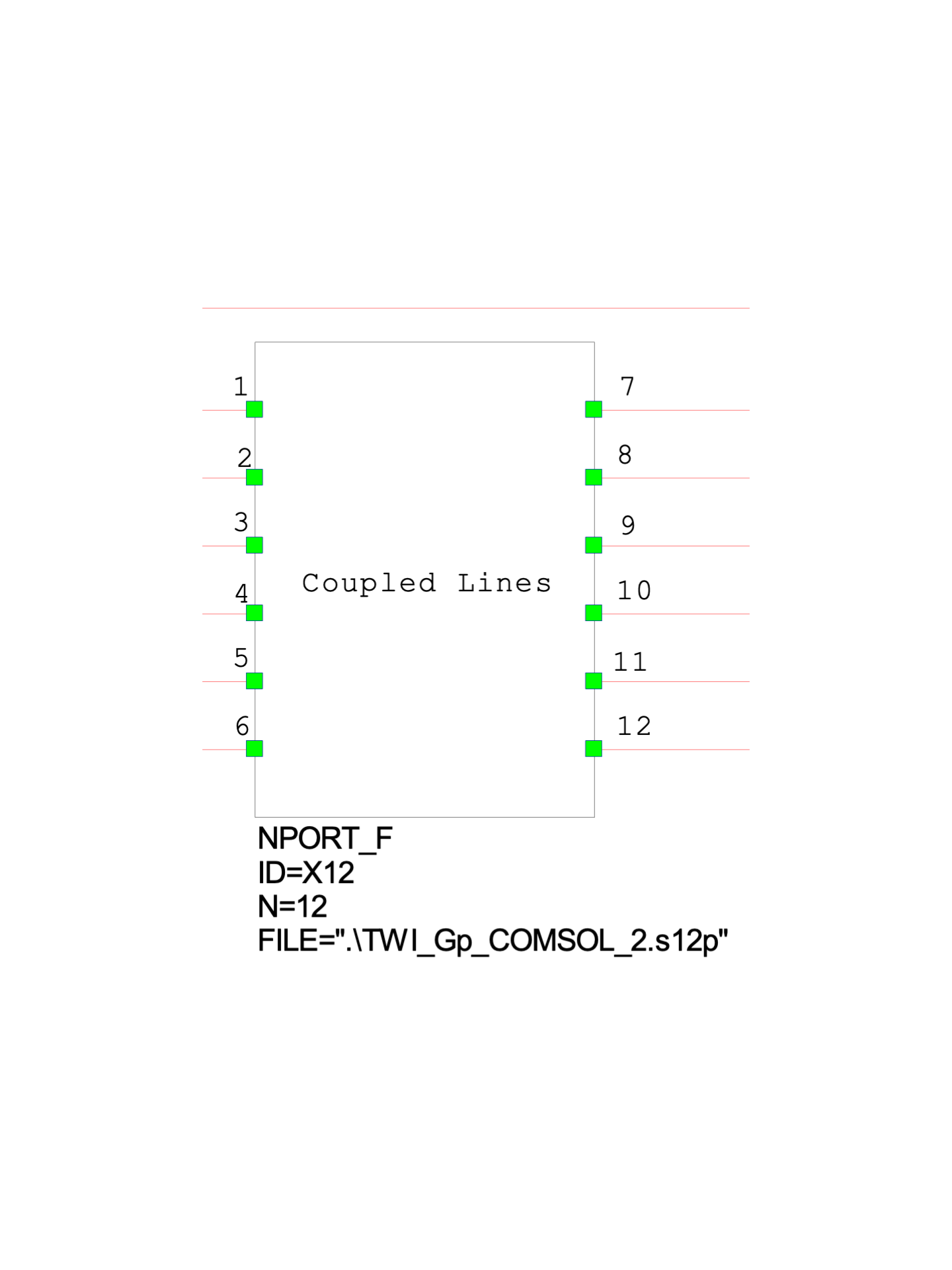}
    \caption{The system schematic of the inverse-designed propagator module (metastructure), implemented in AWR Microwave Office\textsuperscript{\textregistered}. Note that the inputs are the ports 1-5 while the outputs are 6-12 (7 outputs in total).}
    \label{swiss}
\end{figure}

\clearpage

\bibliography{bibliography.bib}
\bibliographystyle{ieeetr}